# Probing the Dark Energy
# in the Functional Protein Universe

*Ezequiel A. Galpern[1,2], Carlos Bueno[3], Ignacio E. Sánchez[1,2],*
*Peter G. Wolynes[3*] and Diego U. Ferreiro[1,2*]*

1 - Protein Physiology Lab, Departamento de Química Biológica, Facultad de Ciencias Exactas y Naturales, Universidad de Buenos Aires, Buenos Aires C1428EGA, Argentina.
2- Instituto de Química Biológica de la Facultad de Ciencias Exactas y Naturales, Consejo Nacional de Investigaciones Científicas y Técnicas - Universidad de Buenos Aires, Buenos Aires C1428EGA, Argentina
3 - Center for Theoretical Biological Physics, Rice University, Houston, TX 77005, USA.
*Corresponding authors: pwolynes@rice.edu, ferreiro@qb.fcen.uba.ar

**Abstract**

We show how to localize and quantify the functional evolutionary constraints on natural proteins. The method compares the perturbations caused by local sequence variants to the energetics of the protein folding process and to the corresponding change to the apparent selection landscape of sequences over the evolutionary time scale. The difference between the physical folding free energies and the evolutionary free energies can be called a 'Dark Energy'. We analyse various protein sets and thereby show that Dark Energy is largely localized at functional sites, which are often energetically frustrated from the point of view of folding. Overall, we find that about 25% of the positions of the folded globular proteins display some significant Dark Energy. When a function relies on a free energy that can be thermodynamically quantified, such as a binding energy to a partner, the relationship of this physical free energy with Dark Energy can be used to define a Functional Selection Temperature. We show that selection for folding and binding functions bear similar weights in specific protein-protein interactions.

# Introduction

Together with the nucleic acids, natural protein molecules are the fundamental entities upon which Life depends. Unlike polymers of amino acids with random sequences, many evolved proteins fold upon themselves to organized structures having some geometrical specificity. The information required not only to specify a protein structure and to encode its distinct activities has been encoded in its sequence through a long process of mutational drift constrained by natural selection. In the fullness of time it appears a fair sample of possible sequences that satisfy the needs of selection has been produced [1]. Using statistical tools to characterize the resulting ensemble of sequences found in Nature thus promises to give insights into selection constraints and in how these constraints are realized physically.

The necessity of proteins to fold to somewhat precise structural ensembles is needed for many functions, although some proteins remain disordered during part of their *lives*. Folding is possible for sequences that give energy landscapes that satisfy the 'Principle of Minimal Frustration' [2]. This principle holds that as a protein folds its free energy in its environment decreases progressively more than would be expected by chance. This energetic bias to the native state overcomes both the roughness of the landscape which would inhibit search for most complex sequences as well as the entropic cost of chain folding. By ensuring that the native basin is a dominant global free energy minimum of a landscape made up of many cooperating interactions we see the minimal frustration principle leads to a landscape that resembles a rough funnel [3]. In parallel to this ontogeny of folded protein structures, the phylogeny of proteins has allowed the exploration of the sequence landscape in search for significantly high fitness values reflecting how well each protein performs functions within an individual organism in a given environment [4]. If the evolutionary landscape of fitnesses were very rugged presumably functional proteins could not have evolved. The evolutionary landscape must contain accessible pathways allowing them both for structural search and fitness improvements [5]. The huge amount of sequence data available today allows us to quantitatively explore the relationships between the structural landscape for folding and the functional landscapes.

Under the assumption of the dominance of foldability as a selection constraint, one can show a quantitative connection between the folding landscape of ontogeny and the evolutionary landscape of phylogeny, using the statistical mechanics of spin glasses and neural networks [5], [6], [7]. Several studies have inferred, from natural sequence co-variation, that the changes in the evolutionary landscape and folding free energy do largely correspond to each other. This correspondence is not perfect however [8], [9], [10], [11]. An obvious deviation occurs in enzyme active sites, where the necessity of conserving the catalytic amino acids largely arises from the rules of chemical mechanism. Folding of the protein is still needed to allow the catalytic residues to fulfill their function and this supportive function is provided by the rest of the sequence. Often achieving the right chemistry is detrimental to folding stability [12]. The genetic code gives protein sequences a limited albeit large coding capacity, so the requirements for function often conflict with the specification for stable structure [13]. Consequently, local discrepancies with the principle of minimal frustration clearly occur and have been shown to highlight protein regions where functional constraints have been particularly strong during evolutionary history [14], [15]. Now, high throughput novel techniques to measure folding stability at a mega-scale makes it possible to carry out nearly complete deep mutational scans at individual sites to localize and score the contributions of individual residues to folding stability [16]. Machine learning based energy models also allow one to score variant fitness [17], [18], and clever ways of analyzing experiments have been proposed [19], [20]. Here we show how combining these high throughput techniques with the statistical tools of energy landscape theory provides a robust quantitative measure of functional constraints for natural proteins.

The collection of all the possible protein folds can justly be called the Protein Universe [21]. It has been argued that a fair sample of this universe has been explored by the biosphere, accounting for general ~$10^5$ [22] to ~$10^8$ [23] distinct types of folds and myriads of functions. Each of these folds attracts a distinct set of sequences that account for the different 'protein families' [4], [24]. Within each structural fold, many different chemical activities have become evolutionarily specified by tweaking the sequences largely locally. Today, we can scan these sequences and distinguish those regions from those that account for folding energetics. The cosmologists define any discrepancies with the present dominant paradigm as 'dark matter' or 'dark energy'. Similarly, the "dark" regions of the protein universe are those that are not well-characterized by existing experimental or computational methods [25], [26]. In sympathy with them, we define here the discrepancy between the folding landscape and the evolutionary landscape inferred from sequence data as 'Dark Energy'. We present here a heuristic to localize and quantify the features of the landscape that are needed to satisfy the functional evolutionary constraints beyond folding by analyzing the sensitivity of both the energetics of protein folding and those effects of site changes to the evolutionary ensemble. This difference, once put in energetic terms by comparing the physical folding landscape and the evolutionary landscape as done by Morcos et al. [7] allows us to measure the 'Dark Energy' of protein evolution, quantifying the strength of functional constraints.

**Conceptual Framework**

The large number of protein sequences, now available for diverse natural variants sharing common folds, provide sufficient data to use a statistical approach to describe how proteins have evolved. Assuming that over the course of evolutionary history of a protein family a fair sample of the selectable sequence space has been achieved with respect to variant fixation in the population, we can characterize the probability of a sequence $\sigma$ using a Boltzmann-like distribution [27],

$$p(\sigma) = \frac{e^{-\Psi^{evo}(\sigma)}}{Z} \qquad (1)$$

In statistical physics, the dimensionless potential $\Psi^{evo}(\sigma)$ should strictly speaking be termed a Massieu Function or 'Free Entropy' [28], in bioinformatics discussions this quantity is commonly referred to as a Statistical or Evolutionary Energy by analogy with Boltzmann's law for systems where dynamics arises from a Hamiltonian [29], [30]. A good approximation to $\Psi^{evo}$ can be inferred from protein sequence data, through a variety of inversion strategies, the earliest and most intuitive being Direct Coupling Analysis (DCA) [31], [32]. $\Psi^{evo}$ takes the form of a Potts Hamiltonian, with the variables being the amino acid identities found at specific sequence positions. The same idea has been implemented using other machine learning methods such as Variational Autoencoders [33], Transformers [34], [35] or Large Language Models [18]. These other methods do not commonly provide an explicit functional form of $\Psi^{evo}$, but can efficiently compute the changes in probability when the sequence is varied and thus changes in $\Psi^{evo}$.

While most random sequences do not fold, natural protein sequences have the remarkable property of being able to fold into robust and stable structures. Energy landscape theory shows this occurs when the folding temperature $T_f$ of a protein is high compared to the glass transition temperature of random globules $T_g$. Equivalently, this means the free energies of the stable structures for a foldable sequence are well separated by a gap from other deep but random competing structures. We see then that evolution has led to minimally frustrated heteropolymers, where amino acid sequences encode mostly energetically favorable interactions so that during folding, Brownian motion can overcome energetic traps and avoid forming misfolded states. If the evolutionary sequence statistics were to reflect only the necessity to fold, $p(\sigma)$ could also be interpreted as the probability that a given sequence in a protein family successfully yields a folded state having a sufficiently low physical Folding Energy in that structure $E^{fold}(\sigma)$. Owing to the high dimensionality of sequence space we expect sequences that can fold to a specific structure to follow a Boltzmann distribution at an apparent selection temperature $T_{sel}$,

$$p(\sigma) = \frac{e^{-\frac{E^{fold}(\sigma)}{k_B T_{sel}}}}{Z} \quad (2)$$

where $k_B$ is the Boltzmann constant. To encode the minimal frustration principle of Energy Landscape Theory, $E^{fold}$ should be measured as the energy difference between the native state and the average compact misfolded states. The energy statistics of misfolded structures are expected for a fixed sequence to mirror the energy statistics of jumbling the sequence. Under this approximation (which has its limitations [36], [37]), the two Boltzmann factors for foldable sequences and structures match so that to first approximation, the Evolutionary Potential or Massieu Function is nothing more than a rescaled Folding Energy at $T_{sel}$ [6], [7], [29], [30], [38]

$$\Psi^{evo}(\sigma) = \frac{E^{fold}(\sigma)}{k_B T_{sel}} \quad (3)$$

This equivalence has allowed the estimation of the apparent selection temperature $T_{sel}$ for diverse protein families [7], [30], [38], [39]. This temperature quantifies in each case the strength of folding stability as an evolutionary force. Within the simplest approximations of energy landscape theory, based on using the random energy model for both landscapes, $T_{sel}$, $T_f$ and $T_g$ are related to each other

$$\frac{2}{T_{sel} T_f} = \frac{1}{T_f^2} + \frac{1}{T_g^2} \quad (4)$$

When $T_{sel}$ is smaller than $T_f$, so is $T_g$ and the landscape is funnel-like.

Folding, while often necessary, however is not sufficient to perform all biological functions [40]. Because of the finite coding capacity of amino acid chains, coding for function often conflicts at least locally with robust folding of globular domains [41]. Energy landscape theory provides a way of looking for local frustration by seeing whether the local landscape of parts of the protein are funnel-like or need the support of the rest of the protein. Implementing this idea has led to frustration analysis, which has proved useful to localize allosteric hinge residues [42], binding and catalytic sites [14] and fuzzy regions of proteins [43]. Since protein sequences can be pictured as populating a funneled evolutionary energy landscape, with the naturally selected sequences at the bottom and random sequences at the top, we can more directly probe the functional selection pressure on proteins by comparing the folding and the sequence energy landscapes after appropriate scaling with the selection temperatures. These two landscapes, while globally close, disagree locally. We can define the difference between the physical folding energy and the free evolutionary energy as 'Dark Energy' (Fig. 1). For a given sequence $\sigma$, the Protein Dark Energy is taken to be the difference between the folding energy gap and the evolutionary gap, scaled by the folding selection temperature,

$$E^{dark}(\sigma) = k_B T_{sel}^{fold} \Psi^{evo}(\sigma) - E^{fold}(\sigma). \tag{5}$$

The Dark Energy then summarizes in a quantitative way the additional functional constraints beyond folding. Sometimes, these additional constraints may themselves be quantified in pure physicochemical energetic terms, for example when specifying a binding energy to a partner. In this case, we can write using this Functional Energy the full distribution as

$$p(\sigma) = \frac{e^{-[E^{fold}(\sigma)/k_B T_{sel}^{fold} + E^{func}(\sigma)/k_B T_{sel}^{func}]}}{Z}. \tag{6}$$

In such cases then we can define a Functional Selection Temperature $T_{sel}^{func}$. This allows us to compare the evolutionary constraint for proper binding to that for proper folding

$$T_{sel}^{func} = T_{sel}^{fold} \frac{E^{func}}{E^{dark}} \tag{7}$$

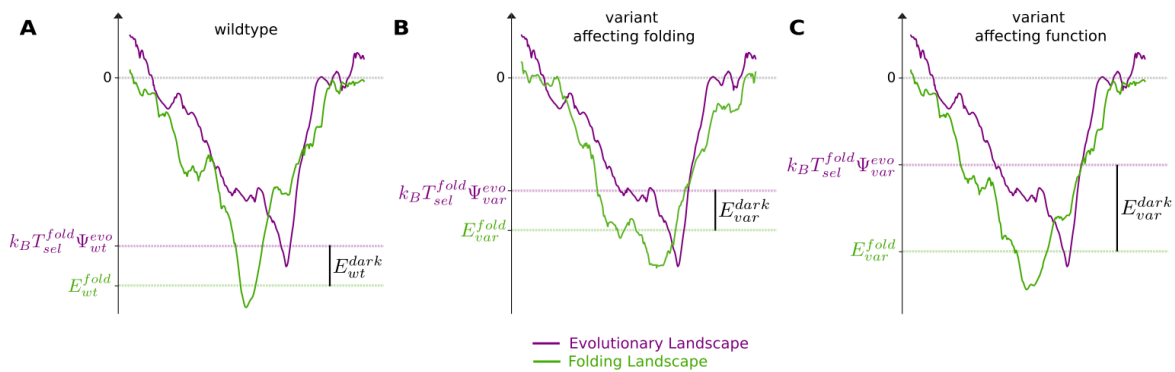

*Figure 1. Dark Energy definition. Schemes of the folding (green) and evolutionary (purple) energy landscapes for a wildtype protein (A), a variant affecting primarily protein folding (B) and a variant affecting an additional function beyond folding (C). The evolutionary landscape is rescaled to be*

*compared with the folding landscape, using the folding selection temperature. The Dark Energy is the difference between the energy gaps of the two landscapes. The folding energy gap is given in each case by the difference between the native state close to the minimum of the funnel and the compact misfolded configurations on the top. In the sequence space, the gap is the difference between a random sequence on top and the evolved sequence close to the bottom. The sequence variant of panel (B) introduces an increase in the folding native energy state that is proportional to the shift in the evolutionary potential, preserving the wild-type Dark Energy. In panel (C), the variant occupies a higher energy state in the evolutionary landscape, but the folding landscape is not affected that much, then Dark Energy is higher for this latest variant.*

We highlight that, although strictly $E^{dark}$ is a Free Energy, we choose to call it just 'Dark Energy' to simplify our language. Given the definition of equation 5, It is particularly of interest to measure the changes in Dark Energy between a wildtype sequence $\sigma_{wt}$ and a variant made at a single site $\sigma_{var}$,

$$\Delta E^{dark}(\sigma_{wt}, \sigma_{var}) = k_B T_{sel}^{fold}[\Psi^{evo}(\sigma_{var}) - \Psi^{evo}(\sigma_{wt})] - [E^{fold}(\sigma_{var}) - E^{fold}(\sigma_{wt})].$$ (8)

Most of the time, single site variations from a natural protein sequence $\sigma_{var}$ will be deleterious. Often this impact on the sequence probability is due to desestabilization of the native structure. In this case, we do not anticipate a Dark Energy change between the wild-type and the variant (Fig. 1B). However, if the variation in the sequence affects a protein function beyond folding, a significant Dark Energy change is expected (Fig. 1C).

The changes in the corresponding folding energies $\Delta E^{fold}$ can be found using experimentally obtained $\Delta\Delta G$ values when available. Such results from a Deep Mutational Scanning are now available in Rocklin's Mega-Scale Dataset [16]. Alternatively, the changes $\Delta E^{fold}$ can also be approximated computationally using a physical force-field, such as the Associative memory, Water mediated, Structure and Energy Model (AWSEM) [44] if experimental values are not yet known. AWSEM is particularly well suited for finding physical folding energies, because it is a coarse-grained and transferable potential that also has been extensively used in Frustration Analysis to localize functional regions from the protein structure and sequence. $\Delta\Psi^{evo}$ can either be computed using the Potts model obtained from DCA or by computing the changes using a protein large language model such as ESM-2 [18], that can be applied to individual sequences without any retraining. In the following sections, we will focus on systematically quantifying the functional effects of protein single-site variants by computing the Dark Energy changes. In addition, for protein-protein interactions systems we will leverage Binding Free Energy experimental data to estimate Functional Selection Temperatures for binding.

## Results

**Protein Dark Energy using experimental folding stability data**

We first made a survey of Dark Energy changes resulting from making single-sites variants, $\Delta E^{dark}$ for 64 proteins studied in a recent Mega-scale folding stability experiment [16] which are available at the ProteinGym database [45]. We used these experimental mutational scannings of $\Delta\Delta G$ to measure the folding energy gap changes, and used the pLM ESM-2 scores [18] as evolutionary free energy changes (see Methods). To illustrate the idea, we show the results for these two quantities for the N-terminal domain of the human FK506-binding protein (PDB: 2KFV), in Fig. 2. To compute the Dark Energy from equation Equation 8 we also need the folding selection temperature $T_{sel}^{fold}$ for each protein family. We can compute this temperature using the same data, by taking a ratio of the two standard deviations of the effect of variations on the two energies from site mutations over the whole protein, $k_B T_{sel}^{fold} = sd(\Delta\Delta G)/sd(\Delta\Psi^{evo})$. This allows us to rescale the evolutionary scores to physical folding energy units. This way of estimating $T_{sel}^{fold}$ has been previously used to estimate the Selection Temperature [30], [39]. The folding selection temperature determined from the dataset gives an average of $T_{sel}^{fold} = (156 \pm 9)K$ (Fig. S1), which being below $T_f$ clearly satisfy the minimal frustration principle and is within the expected theoretical range [46].

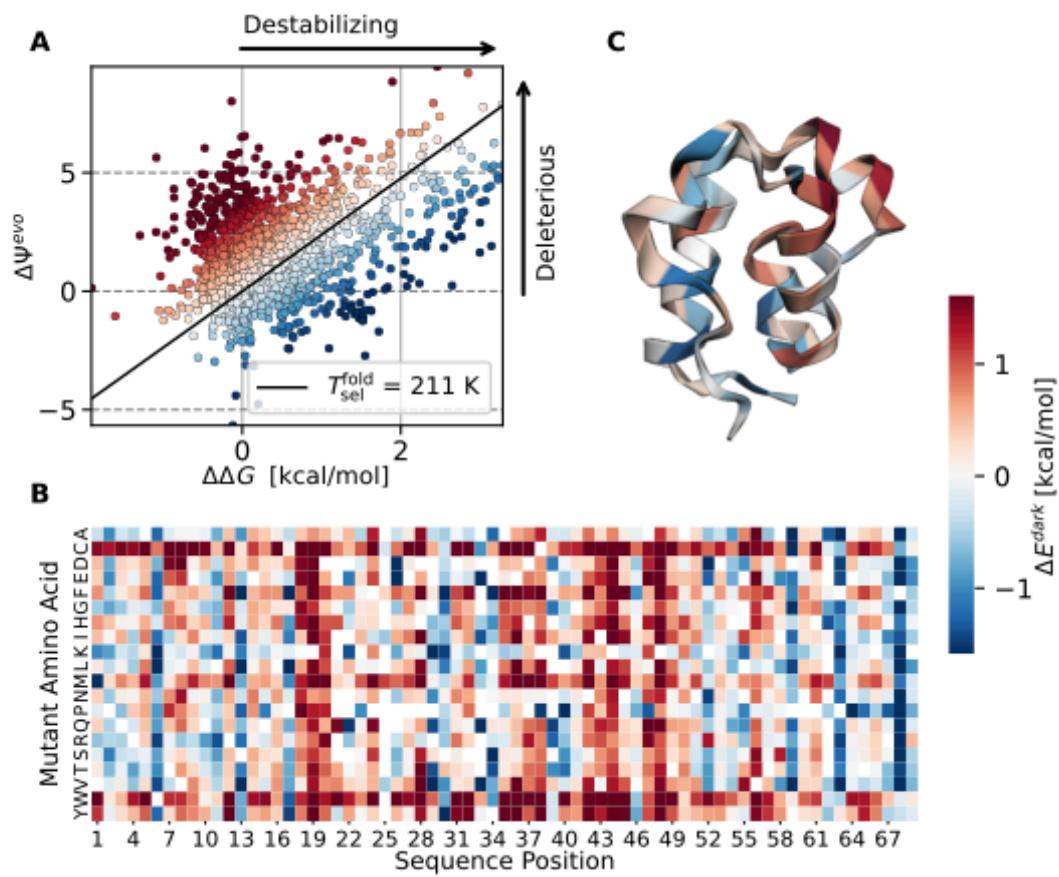

*Figure 2. **Dark Energy deep mutational scanning of a small two-state protein domain (PDB: 2KFV).** (A) ESM-2 evolutionary scores vs folding stability measurements for all single-site variants. The black line slope corresponds to the calculated folding selection temperature for this system. The color scale is given by the Dark Energy change between the wildtype and each variant. (B) A mutational heatmap displaying the same data as in panel (A). High dark energy changes (red) concentrate in some sequence positions. (C) Tertiary structure of the wildtype protein coloured by the weighted site average dark energy change. Regions of high dark energy preferentially localize at proteins' functional sites.*

Fig. 2A displays all the inferred single-site dark energy variations. $\Delta E^{dark}$ is shown in a color scale for all the single-site variants of 2KFV. While most of the variant dark energy changes seem to be distributed around the line expected from the folding selection temperature (black), one notices there are clear outliers, shown as red points. These outliers represent functionally deleterious mutations that do not proportionally affect folding stability (as the variant in Fig. 1C). High $\Delta E^{dark}$ values indicate a strongly deleterious functional effect. A mutational heatmap (Fig. 2B) displaying the same data now separated out by position in the sequence reveals that high $\Delta E^{dark}$ (red) variants appear in a consistent way for some particular sequence sites and then only for some particular amino acid substitutions. Some variants presenting highly negative $\triangle E^{dark}$ (blue) are also concentrated in particular sites. We localize the sites with high Dark Energy by coloring the tertiary structure with a weighted average of $\triangle E^{dark}$. The average of site sequence variations, here, is averaged using the protein amino acid frequencies to weight the possible substitutions. Interestingly, in the 2KFV system used here for illustration, the high Dark Energy sites turn out to be spatially close to each other in a region that has been previously reported to display significant changes in chemical shift upon DNA binding [47], implicating them in function and also have been identified as functional using a different computational approach [16].

Fig. 3 displays in a similar way several small other structures of two-state proteins, that are likewise coloured by the site-averaged $\triangle E^{dark}$. In these representations, the Dark Energy scale is the same for all the systems, allowing for a clear visual inspection. We see that all of these natural proteins show red positions that, according to our model, correspond to functionally constrained regions rather than positions merely constrained by global folding. These sites of high Dark Energy are not limited to being found in any secondary structure. They appear both in $\alpha$, $\beta$ or even unstructured regions. In addition, generally mutations to Alanine are identified as the most representative of the described weighted site-averages $\triangle E^{dark}$ (Fig. S2). This suggests that lower throughput alanine scanning mutagenesis could be used as a good proxy to fully exhaustive mutation analysis to calculate the dark energy landscape of a protein.

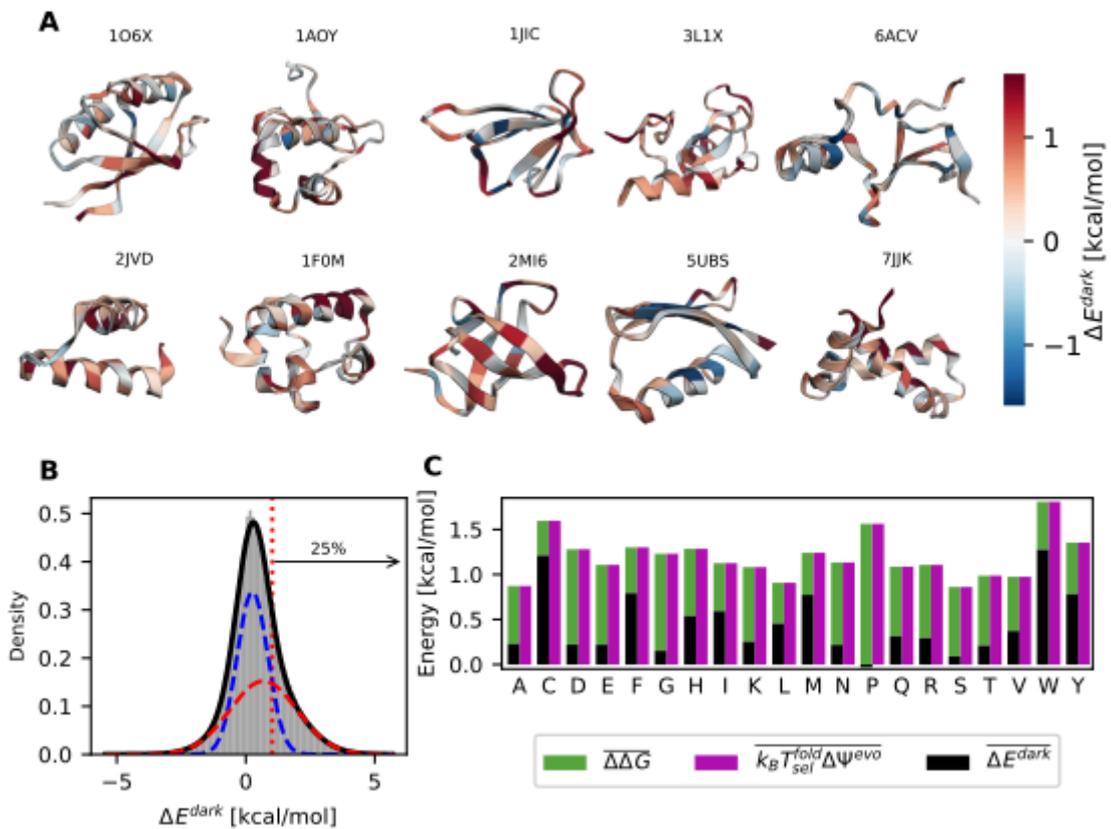

*Figure 3. Dark Energy survey using experimental folding stability data. (A) Tertiary structures of 10 protein domains of the studied dataset, coloured by the dark energy change site-average. (B) Distribution of the Dark Energy changes for all the single-site variants in the dataset weighted by amino acid abundance (gray histogram), The distribution has a heavy right tail (Skewness: 0.273, Excess Kurtosis: 1.220). We fit a Gaussian mixture model (black curve) with two components. The majority of the population is explained by a first component (blue), centered closer to zero than the second component (red). We draw a threshold at 1.03 kcal/mol (red dashed vertical line) where belonging to any of the two distributions is equally probable. The proportion of variants above the threshold is indicated. (C) Folding, Evolutionary and Dark Energy changes averaged for each mutant amino acid.*

In order to get an idea of whether the single-site variants affect primarily stability or are mainly functionally related, we statistically analyzed the dataset overall (Fig. 3B). We find that the distribution of the $\Delta E^{dark}$ for all variants weighted by the amino acid abundance is well described with a two-component Gaussian mixture model (see Methods). The principal component is centered close to zero (blue) while a second component (red) explains most of the high dark energy variants. There is a population with a Dark Energy Change larger than 1.03 kcal/mol where the probability of belonging to the second component is larger than 50%. We propose to use this threshold to separate those variants that affect mainly folding stability (74.7% of this dataset) from the variants that affect mainly protein function (25.3%).

The pattern of effects of amino acid substitution seen in the specific case study, 2KFV, (we have already discussed horizontal dark red stripes in Fig. 2B) is seen also over the dataset. Mutations to Cysteine (C) and Tryptophan (W) consistently exhibit the highest Dark Energy variation (Fig. 3C). This trend correlates with their relative scarcity, as C and W are the least abundant amino acids in the dataset. On average, both the Dark and Evolutionary Energy variations inversely correlate with the relative abundance of the amino acid to which the sequence is changed (Fig. S3). Mutations to Proline (P) are particularly interesting: despite appearing in slightly greater abundance than Tryptophan (W), such mutations present an almost negligible average Dark Energy variation (Fig. 3C). In this case, the large deleterious effect is explained by a proportionally large folding destabilization, aligning with Proline's known role in disrupting secondary structure.

Finally, we have also recalculated for the Rocklin dataset the $\Delta E^{dark}$ using the coarse-grained transferable AWSEM potential [48] to compute the input $\Delta E^{fold}$ (see Methods) for each protein in the dataset rather than the experimental values. We find that $\Delta E^{dark}$ calculated in this way correlates well with the results obtained using experimental $\Delta\Delta G$ (average Pearson's r of 0.63, Fig. S4A). Again modeling the AWSEM-calculated $\Delta E^{dark}$ distribution as a Gaussian mixture with two components (Fig. S4B), we find that the proportion of the variants that seem to affect solely protein function is somewhat larger than was found using the Rocklin $\Delta\Delta G$ values (36.4%) perhaps reflecting inadequacies of the force-field. The weighted site $\Delta E^{dark}$ averages, nevertheless conserve the functional patterns observed when using the actual experimental data (Fig. S4C-E). This concordance allows then the use of computed $\Delta E^{fold}$ values to study easily other proteins outside the Rocklin dataset.

**Dark Energy in enzymes**

We turn our attention now to Dark Energy changes $\Delta E^{dark}$ in enzymes. We study protein enzymes that already have experimentally annotated catalytic sites. We survey the Catalytic Site Atlas (CSA) [49], in this case measuring $\Delta E^{dark}$ using ESM-2 evolutionary scores as $\Delta\Psi^{evo}$ and using the theoretical route through AWSEM to estimate the folding energy gap changes upon sequence variations of specific sites $\Delta E^{fold}$ (see Methods).

To illustrate the results we first highlight the *E. Coli* TEM1 Beta-Lactamase as a case study (Fig. 4A-B). Almost all of its catalytic residues -Ser70, Lys73, Ser130, Glu166 and Lys234- display extremely high dark energy change site averages ($\Delta E^{dark} > 3$ kcal/mol). As a reference, the threshold determined using AWSEM for the dataset studied in the previous section is 1.36 kcal/mol (Fig. S4). Virtually any variation at the highly conserved catalytic residues significantly reduces the evolutionary gap $\Delta \Psi^{evo}$, while the folding gap $\Delta E^{fold}$ remains unchanged or sometimes even is deepened. A visual inspection of the colored structure on Fig. 4A reveals that, besides the described effect at the famous catalytic triad, high Dark Energy changes (red) also are enriched around the active sites, decaying as one moves away from the catalytic center. This pattern of a halo of Dark Energy around the active sites parallels the halo of frustration seen in surveys of local frustration in enzymes [14]. It also is consistent with the observation from studies of artificial protein evolution that often the needed mutations for optimal catalysis are to be found in a second shell around the active site [50].

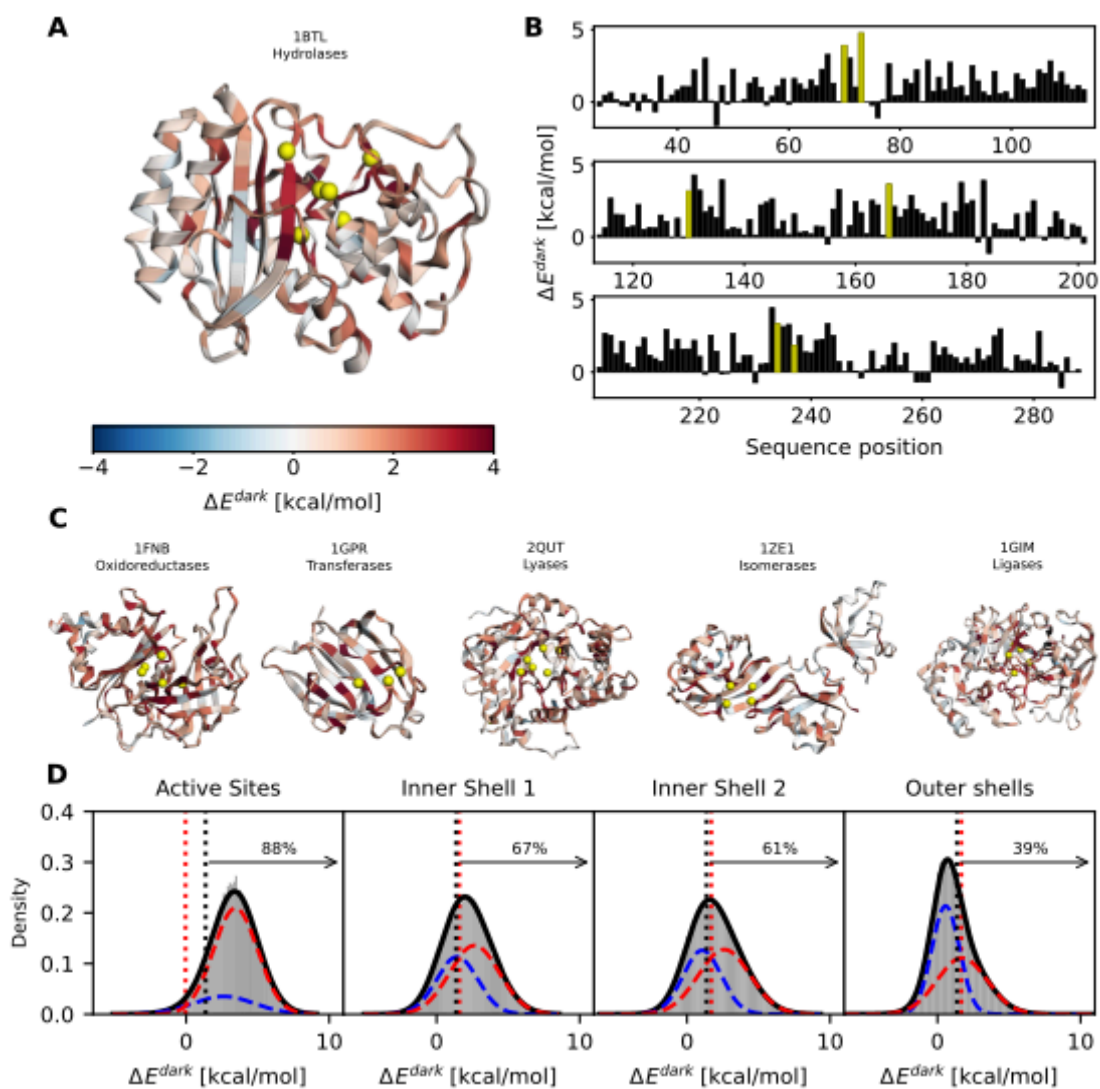

*Figure 4. **Dark Energy changes in protein enzymes.** (A) E. Coli TEM1 Beta-Lactamase structure (1BTL, Hydrolase class) colored by Dark Energy changes weighted site-average using AWSEM and ESM-2. Annotated active sites are highlighted with yellow spheres. (B) For the same protein, Dark Energy changes weighted site-average for each residue, with active sites in yellow. (C) An example protein structure colored by Dark Energy changes weighted site-average for different enzyme*

*classes. Active sites are marked with yellow spheres. The color scale is shared with panel (A). (D) For all the studied variants in the Catalytic Site Atlas* [49]*, Dark Energy changes weighted histogram (gray) for the active sites, the sites in Inner Shell 1, 2 and the Outer Shells (see Methods). A two-component Gaussian mixture model is shown in a black line, while the components are the blue and red lines. For each case, we draw a threshold (red dashed vertical line) where belonging to any of the two distributions is equally probable. The black dashed vertical line is the threshold of 1.36 kcal/mol obtained for the previous dataset using AWSEM. The weighted percentage of variants above the threshold for each shell is shown.*

The patterns that we described for Beta-Lactamase are not an exception. The same patterns were found across all the enzyme classes of the studied dataset (Fig. 4C). If we now apply a two-component Gaussian Mixture Models to the $\Delta E^{dark}$ weighted distribution first for the active sites, then for the residues in contact with an active site (Inner shell 1), the residues in contact with those (Inner shell 2) and for the rest of the protein (Outer shells), we find that the second component (red in Fig. 4D) smoothly gains weight as the sites get closer to the specific chemically active residues. Using the 1.36 kcal/mol threshold (Fig. S4) for significant function, functional-related variants comprise 38.9% of the variation in the Outer shells, 60.7% of them the Inner shell 2, 66.8% for the Inner shell 1 and 87.6% for the active sites. The remaining fraction of the variants then are mostly folding-related. Again we see that sometimes functional variants appear also far away from the annotated active sites. There, the high $\Delta E^{dark}$ values reveal allosteric hinge regions or binding sites to macromolecular partners.

**Binding Selection Temperature estimation**

When sequence evolution can be viewed to being constrained by folding and a single function that can be quantified through a physicochemical energy, equation Equation 7 allows the calculation of a Functional Selection Temperature $T_{sel}^{func}$ by comparing the Dark Energy to the corresponding Functional Energy, that can be measured in the laboratory or can be computationally determined. This can then be compared to $T_{sel}^{fold}$. A very nice example to apply this framework is the Barstar-Barnase system, because of the extensive binding and folding measurements made in the Fersht laboratory [51], [52]. Barstar is the intracellular specific inhibitor of Barnase, an extracellular ribonuclease of *Bacillus amyloliquefaciens* that becomes lethal inside the cells of other organisms but not in the *Bacillus* itself by virtue of its specific binding to Barstar [51]. It would seem clear that both Barnase and Barstar evolution have been constrained not only by folding requirements but also by the 'binding to Barnase' activity. Barstar Dark Energy changes $\Delta E^{dark}$ weighted site-averages, obtained using AWSEM and ESM-2, are shown in Fig. 5A. Red regions indicate there are stronger functional than folding requirements. They appear primarily in the binding interface to Barnase. The Folding Selection Temperature of Barstar is $T_{sel}^{fold} = 141 K$ . For the 'binding to Barnase' function, the corresponding Functional Energy changes upon single-site mutants are $\Delta \Delta G^{binding}$. These binders effects have been experimentally determined for several variants [52]. We find that, as expected $\Delta \Delta G^{binding}$ strongly correlates with $\Delta E^{dark}$ (Fig. 5B, *r* = 0.87). The slope of this linear relationship corresponds to the ratio of the selection temperatures $T_{sel}^{bind}/T_{sel}^{fold}$, from which we obtain $T_{sel}^{bind} = (186 \pm 31) K$, not terribly far from the folding selection temperature.

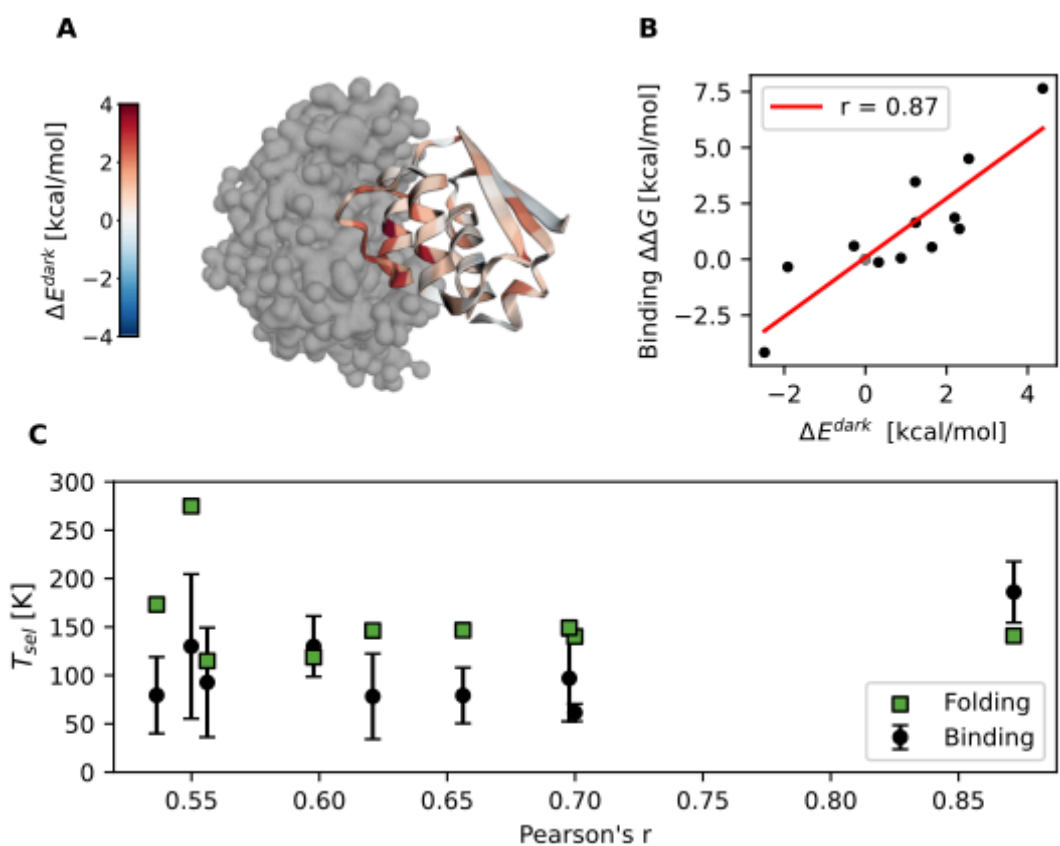

*Figure 5. Binding Selection Temperature. (A) Barstar (ribbon structure) colored by the Dark Energy Variation weighted site-averages, binding to Barnase (PDB: 1BRS, chains A and D). (B) For Barstar, single-site mutant Binding ΔΔG to Barnase as a function of Dark Energy changes and a linear fit to the data (red line). (C) Folding and Binding Selection Temperatures for different Protein-Protein interacting partners as a function of the Pearson's correlation coefficient between Binding ΔΔG and Dark Energy changes (see* Table S1*).*

We can extend this same methodology to a set of Protein-Protein interacting partners for which experimental values of $\Delta\Delta G^{binding}$ have been annotated in the Skempi database [53] (see Methods). We include the results for the 32 studied cases in Table S1. For the cases where the correlation between $\Delta\Delta G^{binding}$ and $\Delta E^{dark}$ is strong enough (Pearson's r > 0.5), we can then compute a Binding Selection Temperature (Fig. 5C). The resulting $T_{sel}^{bind}$ values, ranging from 60 to 186 K, are comparable to $T_{sel}^{fold}$ values, though no consistent relationship has been observed. The fact that some selection temperatures for binding are low suggest they must be strong binders to low concentration partners. Several reasons can explain the lack of correlation between $\Delta\Delta G^{binding}$ and $\Delta E^{dark}$ for some of the studied systems (Table S1). Of course, the idea that one single function beyond folding constrains the evolution of some of these proteins is likely an oversimplification. We also note that the available $\Delta\Delta G^{binding}$ experimental data for these systems are not extensive -only a few site mutants have been tested for each protein-.

**Discussion**

Strictly speaking, the concept of fitness should be applied to populations of evolving organisms through their ability to pass on genes after enduring the rigors of life [54]. To a certain extent in molecular biology, this theoretical object for whole organisms can sometimes be reduced to the selection of 'biological functions' of individual proteins for which structural genes code, or for assemblies of proteins in biological pathways, which are for the most part, foldable sequences of amino acids of functional molecules.

Protein folding turns out to be a major driver in the evolution of coding sequences of functional molecules [55], [56]. Nevertheless, folding is a necessary but not a sufficient condition to carry out the myriad of biological activities that modern proteins mediate. Many of these activities are still uncharacterized or incomplete. In analogy to cosmology, the regions of proteins where molecular conformation is unknown has been referred to as the Dark Matter of the protein universe [26]. Here we have put forward a general framework based on Dark Energy to quantify the impact of the 'biological functions' on the evolution of protein energetics by analyzing the effect of amino acid variations on the folding and evolution of proteins.

We have shown how to probe the functional selection pressure on proteins by comparing the folding and the evolutionary sequence energy landscapes under appropriate scaling and then defining their difference as the 'Dark Energy' (Fig. 1, Equation 5). For any mutant, the Dark Energy can be computed when we have a good proxy of the evolutionary likelihood of the mutation ($\triangle \Psi^{evo}$) and for the thermodynamic effect of this same substitution on protein folding ($\triangle E^{fold}$) (Equation 8). Dark Energy thus measures evolutionary variations that cannot be attributed to changes in folding energetics alone. We remark that the proposed metric is not just an *ad-hoc* score but provides a common scale that can be used across proteins and has physical energy units.

In principle Dark Energy changes can be null, positive or negative, depending on the effect that a mutation has on either or both of the physical and evolutionary landscapes (white, red or blue in Fig. 2, see distributions in Fig. 3 and Fig. 4). By analyzing an extensive dataset of experimental mutants, we have found that most of the Dark Energy changes are positive, in between 1 and 4 kcal/mol (Fig. 4). These account for about 25% of the total variants. These changes are more deleterious than can be expected by their effect on protein folding alone (white in Fig. 2), and in some cases may even make the variant more stable than the wild type. These mutations likely would result in loss-of-function phenotypes while maintaining protein stability, indicating a functional rather than a structural limitation. There is however a minority of the possible single-site variants that do induce a negative change in Dark Energy, overstabilizing the protein. We note that the magnitude of the Dark Energy changes is on the order of the changes that can be introduced by amino acids variations on the folding energetics, making $T_{sel}^{fold}$ within the expected theoretical range being much lower than the physiological folding temperature (Fig. S1). This supports the quantitative treatment of the minimal frustration principle by Bryngelson and Wolynes [2]. The functional effects of mutations can be easily seen in their structural context because their magnitude is comparable to the changes introduced by sequence variations in the folding energetics. In agreement with previous work [1], [15], the present analysis shows that there is often a trade-off between function and folding, where the fixation of functional sites compromises folding. This is a consequence of there being only a finite coding alphabet. It has been suggested that most natural protein sequences turn out to be close to saturating their information capacity coding needed for folding [13]. Folding is encoded globally across the structure, while coding for function can be done locally. Due to this localization, the regions that are identified as enriched in Dark Energy are not randomly distributed over the structures but largely cluster together (Fig. 2, Fig. 3, Fig. 4, Fig. 5). Taking all the proteins together, we see that the positions with high weighted site average Dark Energy make up as much as 25% ~ 35% of the proteins, depending on the method used to infer $\Delta E^{fold}$. This fraction does not appear to vary much with the protein size.

First by examining several anecdotal examples, we have shown that Dark Energy tends to localize in known functional regions of proteins (Fig. 2, Fig. 4, Fig. 5). In a statistical survey of enzymes we found that Dark Energy localizes around the catalytic centers and then smoothly decays as one leaves the center's vicinity (Fig. 4). This is consistent with the frustration analysis that shows that catalytic sites are enriched in interactions that conflict with robust folding of the domains [14]. Interestingly, there are sometimes sites of high Dark Energy far away from the catalytic sites, marking places where mutations can be expected to affect function in other ways. Presumably these sites contribute to important enzyme dynamics and allostery [50], [57], [58].

When sequence evolution has been constrained by folding and only a single other 'biological function', which can also be quantified through a physicochemical energy, we can calculate a Functional Selection Temperature $T_{sel}^{func}$, using Equation 7. We analyzed a dataset of protein-protein interactions for which the effects of mutations on binding affinity have been experimentally measured, and assigned these changes in functional free energy as such a physicochemical 'energy'. For the archetypical case of the Barstar-Barnase pair, we find that the functional binding energy nicely correlates with the Dark Energy (Fig. 5). The slope of this relation provides the ratio $T_{sel}^{func}/T_{sel}^{fold}$, which in this case is close to unity. This indicates that most of the 'biological function' of Barstar can be accounted for by its binding to Barnase. In general we found that $T_{sel}^{func}$ is on the order of $T_{sel}^{fold}$, suggesting that the strength of the sequence restrictions for binding in functional sites are similar to that for folding. We found however that only in a third of the analyzed systems can the Dark Energy be accounted entirely for by their 'binding function' (Table S1).

      The quantification and localization of Dark Energy can provide a test for whether a codable and foldable molecular system is under evolutionary constraints, even when the 'biological function' of the system may be unknown. The impact of functions on the coding sequence leaves a trace in the perturbations of the folding energetics. We expect that whenever the function that exerts the evolutionary pressure disappears, the folding energy needed to satisfy the minimal frustration principle would redistribute along the structure, simply for entropic reasons, and so over evolutionary time the Dark Energy for such external function will tend to disappear [38].

**Methods**

**Folding Stability Experimental Data.** We used the subset of 64 deep mutational scanning assays of the 'Mega-scale experimental analysis of protein folding stability' [16] available at the ProteinGym database [45]. We used the thermodynamic folding stability measurements (ΔΔG) and the correspondent ESM-2 (33 layers and 650M parameter version) Zero-Shot scores [18] for single-site substitutions. We impose the sign convention where ΔΔG<0 for stabilizing variants.

**Active Sites Data.** We studied the PDBs for all the entries of the Catalytic Site Atlas [49]. We discarded those structures with any sequence mismatch between the PDB sequence and the annotated sites, keeping a total of 732 PDBs.

**Binding Energy Data.** We selected from the Skempi database [53] the entries of proteins interacting with a single protein partner with 5 or more single-site mutants. We converted the affinities of the wild-type and mutant to ΔG values using the annotated temperature of the experiments. We impose the sign convention where ΔΔG<0 for variants that stabilize the complex.

**Evolutionary Score.** We estimated $\Delta \Psi^{evo}$ by computing the Evolutionary Score of the the 33-layer, 650-million parameter version of Evolutionary Scale Modeling-2 (ESM-2), trained on the UR50D dataset [18]. For a given sequence $\sigma = (\sigma_1, \ldots, \sigma_L)$, the score for each possible single-residue substitution is calculated by masking each position independently and computing the logits with the model. The logits $l_i$ are the unnormalized log-probabilities used by the model to estimate the conditional distribution over amino acids at the masked position. For a mutation $\sigma_i = a \to b$, the evolutionary score is a log-likelihood ratio between the mutant and the wildtype residues, according to the model:

$$\Delta \Psi^{evo} = -[l_i(b) - l_i(a)] = -log\left(\frac{P(b \mid \sigma_{\setminus i})}{P(a \mid \sigma_{\setminus i})}\right)$$

As a sign convention, higher scores indicate that the mutation is less likely under the model. We computed the logits using a custom script that integrates Python functions adapted from [59].

**Computational estimation of folding stability changes.** We calculated the folding energy of protein structures using the transferable AWSEM coarse-grained force field [44], including only the burial and the contact terms. For each single-site variant, we compute the energy difference given by the sequence change in the forcefield, assuming the tertiary native structure is conserved. We rescale the AWSEM energy values to experimental folding energy units, using as a reference the 64 studied proteins of the Mega-Scale dataset. For each, we compute a scaling factor given by the ratio of standard deviations $sd(\Delta\Delta G)/sd(\Delta E^{AWSEM})$. We get on average that 1 AWSEM unit = 0.06 kcal/mol.

**Gaussian Mixture Model.** We fitted the Dark Energy Variation Weighted distributions with two-component Gaussian Mixture Models. The weights are given by the amino acid abundance of each dataset. We establish the pertinence of a two-component model by measuring Skewness and Excess Kurtosis in each case, obtaining positive values in all cases, and comparing the chi-squared and Akaike Information Criterion (AIC) values. We fit the data using the python library *sklearn*, testing multiple initial conditions to ensure a robust minimization of the chi-squared.

**Enzyme Shells.** To analyze the enzymes, we classify its sequence sites into four Shells: Active Sites, Inner Shell 1, Inner Shell 2 and Outer Shells. The annotated active sites belong to the Active Site Shell. The residues at less than 5A from the active sites that are not annotated as catalytic themselves belong to the Inner Shell 1. The residues at less than 5A from the ones in the Inner Shell 1 that do not belong to the previous shell are classified as the Inner Shell 2. The remaining residues belong to the Outer Shells.

**Data Availability**

All the protein sequences and PDB IDs used in this work are hosted in the linked Zenodo repository [doi.org/10.5281/zenodo.16413074](doi.org/10.5281/zenodo.16413074).

**Code Availability**

A python package that allows for the calculations of Dark Energy and Folding Energy estimations using AWSEM is available at [github.com/HanaJaafari/Frustratometer](github.com/HanaJaafari/Frustratometer). A Google Colaboratory Notebook for computing and visualizing Dark Energy changes for any PDB structure is available at
[colab.research.google.com/github/eagalpern/colabs/blob/main/DarkEnergy.ipynb](colab.research.google.com/github/eagalpern/colabs/blob/main/DarkEnergy.ipynb)

**Acknowledgments**

We thank Mafalda Dias and Jonathan Frazer for insightful discussions. DUF recognizes T.H.E.T.A. (Tradicional Hermandad Empeñada en la Transformación del Ábaco) for unconditional support. EAG, IES and DUF are supported by the Consejo de Investigaciones Científicas y Técnicas (CONICET); CONICET Grant PIP2022-2024—11220210100704CO and Universidad de Buenos Aires grant UBACyT 20020220200106BA. PGW was supported both by the Bullard–Welch Chair at Rice University, grant C-0016, and by the Center for Theoretical Biological Physics sponsored by NSF grant PHY-2019745. We call the attention of the international scientific community about the catastrophic erosion of Argentina's strong scientific tradition due to current funding constraints and the sudden termination of long term policies.

# Supporting Information for

**Probing the Dark Energy in the Functional Protein Universe**

**Table S1**

**Figures S1 to S4**

| System | Binding Tsel [K] | Error Binding Tsel [K] | Folding Tsel [K] | r | Slope |
|---|---|---|---|---|---|
| Barstar + Barnase | 186 | 32 | 141 | 0.87 | 1.32 |
| RalGDS-RBD + H-Ras1 | 61 | 9 | 140 | 0.70 | 0.44 |
| Cytochrome C peroxidase + Non-cognate Cytochrome C | 97 | 45 | 149 | 0.70 | 0.65 |
| DNA repair protein recO + Single-stranded DNA-binding protein | 79 | 29 | 147 | 0.66 | 0.54 |
| Rac-1 + p67phox | 78 | 44 | 146 | 0.62 | 0.54 |
| Barnase + Barstar | 130 | 31 | 119 | 0.60 | 1,09 |
| Cytochrome C + Cytochrome C peroxidase | 93 | 57 | 115 | 0.56 | 0.81 |
| PDE delta + K-Ras | 130 | 75 | 275 | 0.55 | 0.47 |
| Ubiquitin conjugation factor E4 + DSK2 | 79 | 40 | 173 | 0.54 | 0.46 |
| B domain of Protein G + IgG1 Fc | | | 184 | 0.45 | 1.17 |
| Raf-RBD + Rap1a | | | 152 | 0.33 | 0.24 |
| Interferon alpha_beta receptor 1 + Interferon alpha-2 | | | 146 | 0.32 | 0.22 |
| Turkey ovomucoid third domain + Streptomyces griseus proteinase B | | | 110 | 0.27 | 0.78 |
| TEM-1 beta-lactamase + BLIP-II | | | 151 | 0.25 | 0.23 |
| Turkey ovomucoid third domain + Subtilisin Carlsberg | | | 115 | 0.23 | 0.59 |
| TEM-1 beta-lactamase + BLIP | | | 157 | 0.19 | 0.22 |
| Colicin E2 immunity protein + Colicin E9 DNase | | | 305 | 0.18 | 0.81 |
| Turkey ovomucoid third domain + Human leukocyte elastase | | | 120 | 0.15 | 0.39 |
| Colicin E9 immunity protein + Colicin E9 DNase | | | 400 | 0.14 | 0.38 |
| Angiogenin + Ribonuclease inhibitor | | | 131 | 0.13 | 0.13 |
| Membrane-type serine protease 1 + BPTI | | | 150 | 0.10 | 0.09 |
| Interferon alpha_beta receptor 1 + Interferon omega-1 | | | 156 | 0.10 | 0.07 |
| Ribonuclease inhibitor + Angiogenin | | | 170 | 0.08 | 0.11 |
| Ribonuclease inhibitor + RNase A | | | 171 | 0.03 | 0.06 |
| Interferon alpha-2 + Interferon alpha_beta receptor 2 | | | 180 | -0.02 | -0.02 |
| Rad23 + Ubiquitin conjugation factor E4 | | | 166 | -0.06 | -0.09 |
| dHP1 Chromodomain + H3 tail | | | 122 | -0.16 | -0.16 |
| R. flavefaciens Doc1a dockerin + R. flavefaciens ScaB cohesin 3 | | | 128 | -0.20 | -0.15 |
| Interferon alpha_beta receptor 2 + Interferon alpha-2 | | | 212 | -0.20 | -0.22 |
| BLIP + TEM-1 beta-lactamase | | | 163 | -0.28 | -1.03 |
| Interferon omega-1 + Interferon alpha_beta receptor 1 | | | 180 | -0.35 | -0.01 |
| Chemotaxis protein CheA + Chemotaxis protein CheY | | | 367 | -0.40 | -0.23 |

*Table S1.* Linear fit results and Selection Temperatures for all the studied Protein-Protein binding systems, sorted by the correlation coefficient between Dark Energy changes and Binding ΔΔG.

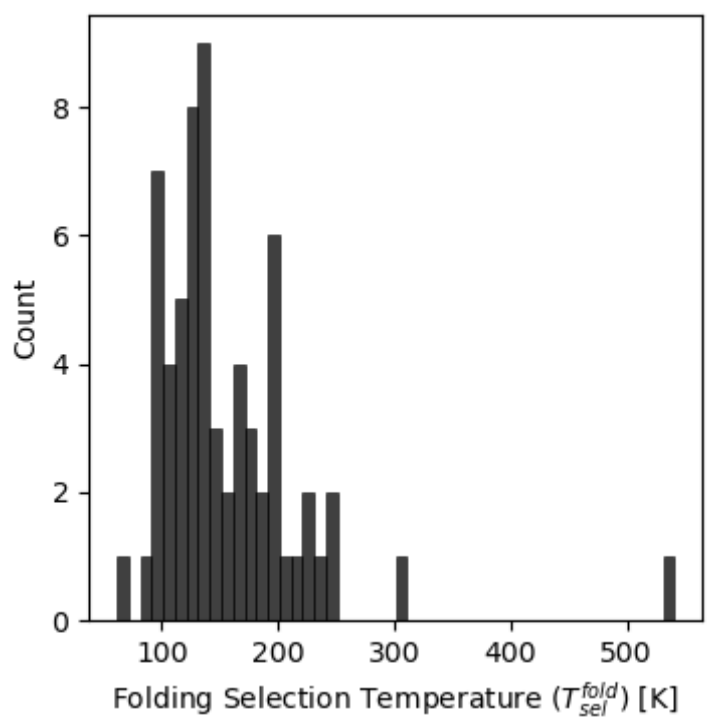

*Figure S1.* Folding Selection Temperature histogram for the 64 proteins with experimental ΔΔG annotations.

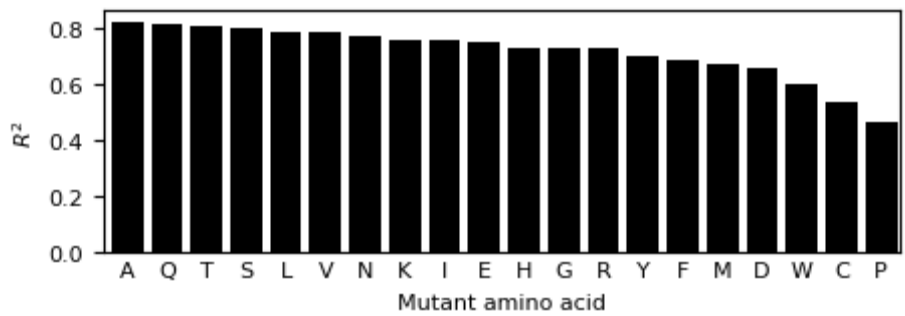

*Figure S2*. Determination coefficient ($R^2$) for the relationship between the Dark Energy change upon mutations to each of the 20 standard amino acids and the weighted site-average of the Dark Energy changes, for the 64 proteins with experimental ΔΔG annotations. Mutations to Alanine (A) show the strongest correlation to the average.

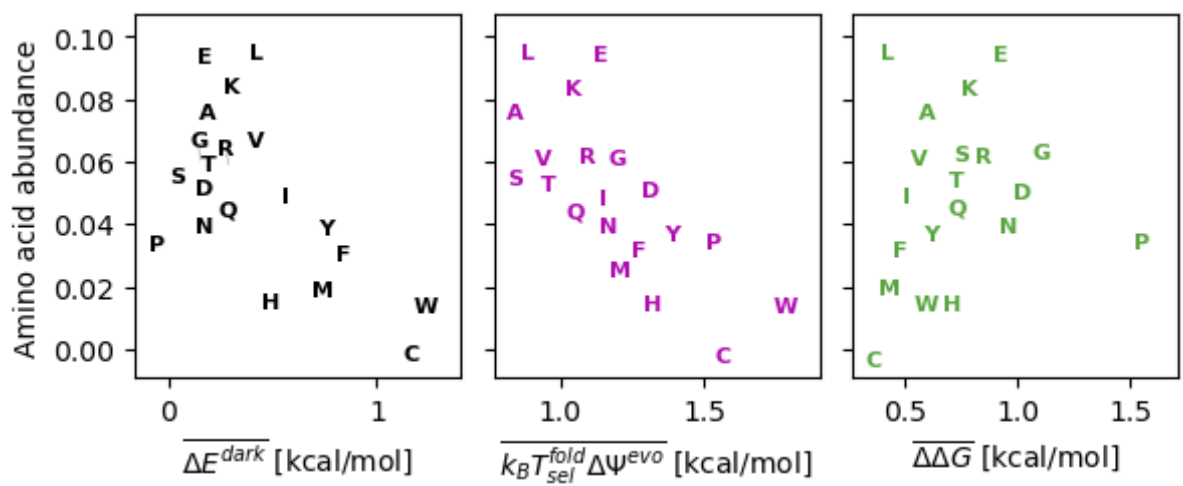

*Figure S3.* Amino acid abundance as a function of Dark, Evolutionary and Folding Energy variations averaged for each mutant amino acid.

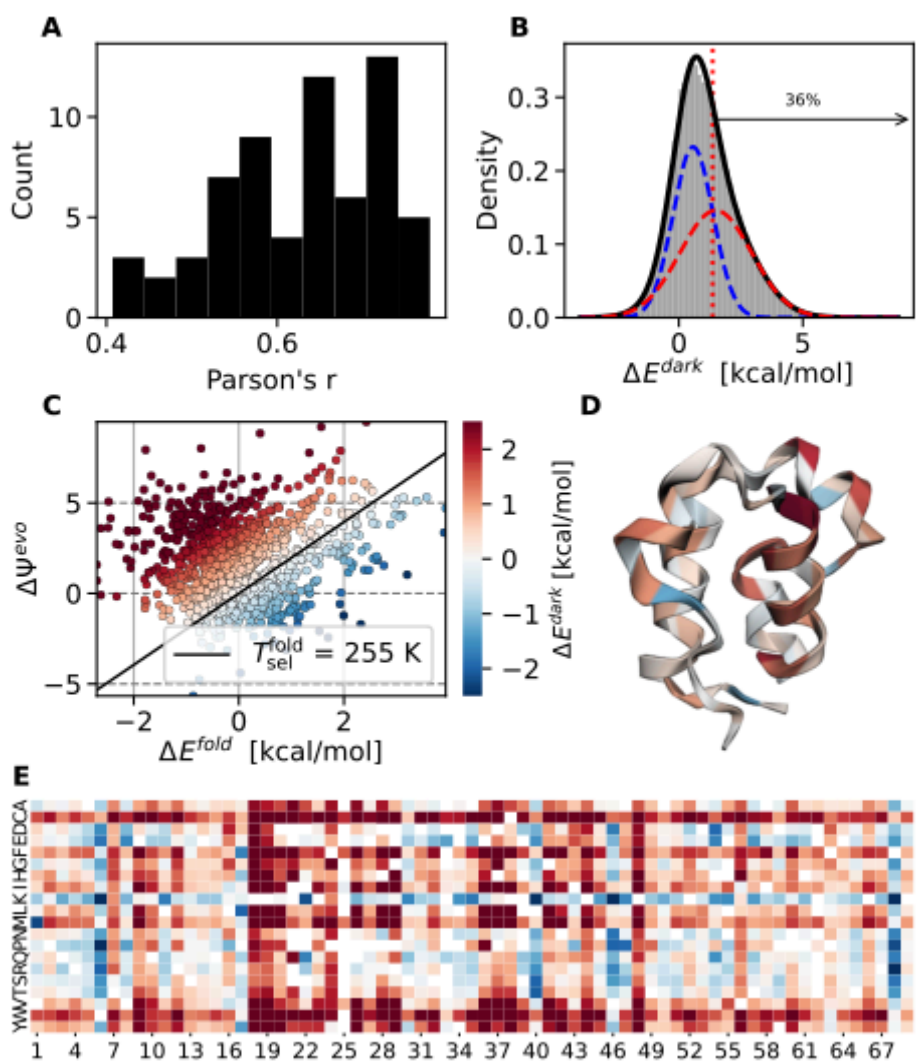

*Figure S4. Dark Energy changes computed with AWSEM.* (A). Histogram of the Pearson correlation coefficients between Dark Energy changes computed using ΔΔG and AWSEM for each of the proteins in the dataset. (B) Distribution of the Dark Energy changes computed with AWSEM (rescaled) for all the single-site mutants in the dataset (gray histogram), weighted by amino acid abundance. The distribution has a heavy right tail (Skewness: 0.452, Kurtosis: 0.532). We fit a Gaussian mixture model (black curve) with two components. The majority of the population is explained by a first component (blue), centered closer to zero than the second component (red). We draw a threshold at 1.36 kcal/mol (red dashed vertical line) where belonging to any of the two distributions is equally probable. The percentage of variants above the threshold is indicated. (C) ESM-2 evolutionary scores vs AWSEM folding energies (rescaled). The black line slope corresponds to the calculated folding selection temperature for this system. The color scale is given by the Dark Energy changes. (D) Tertiary structure of the wildtype protein coloured by the weighted site average dark energy change. (E) A mutational heatmap displaying the same data. High dark energy changes (red) concentrate in some sequence positions.